 \documentclass[preprint,12pt]{elsarticle}
\usepackage{graphicx}
\usepackage{amssymb}
\journal{Journal of Luminescence}

\usepackage{url}
\usepackage[nofiglist,nomarkers]{endfloat}

\begin{document}

\begin{frontmatter}



  \title{Impurity-trapped excitons and electron traps in
    CaF$_2$:Yb$^{2+}$ and SrF$_2$:Yb$^{2+}$ probed by transient
    photoluminescence enhancement}


\author[UC]{P. S. Senanayake}
\author[UC]{J. P. R. Wells}
\author[UC,MD]{M. F. Reid\corref{cor1}} \ead{mike.reid@canterbury.ac.nz}
\author[Felix]{G. Berden}
\author[UU]{A. Meijerink}
\author[UC,MD]{R. J. Reeves}

\address[UC]{Department of Physics and Astronomy, University of
  Canterbury, PB 4800, Christchurch 8140, New Zealand}
\address[MD]{MacDiarmid Institute for Advanced Materials and
  Nanotechnology, University of Canterbury} 
\address[Felix]{FELIX Free Electron Laser Facility, FOM Institute for
  Plasmaphysics Rijnhuizen, PO Box 1207, 3430 BE, Nieuwegein, The
  Netherlands} 
\address[UU]{Debye Institute for NanoMaterials Science, University of
  Utrecht, P.O. Box 80000, TA 3508 Utrecht, The Netherlands}

\cortext[cor1]{Corresponding author. Tel.: +64 3 364 2548;  fax: +64 3 364 2469.}

\begin{abstract}
  CaF$_2$:Yb$^{2+}$ and SrF$_2$:Yb$^{2+}$ crystals have been
  investigated by a two-color UV + IR transient photoluminescence
  enhancement technique. The enhancement gives information about both
  changes in internal energy levels of the excitons and liberation of
  electrons from traps in the crystals.
\end{abstract}


\end{frontmatter}

% \linenumbers

%\clearpage

\section{Introduction}
\label{sec:introduction}

Impurity-trapped excitons in rare-earth doped materials play an
important role in both radiative and non-radiative decay processes of
the excited configurations of the rare-earth ions.
Theoretical~\cite{SaSeBa10a} and experimental~\cite{GrMa08} studies
suggest that non-radiative decay of the excited $4f^{N-1}5d$
configuration to the ground $4f^{N}$ configuration can be mediated by
trapped exciton states, where the $5d$ electron becomes delocalized.
Radiation from exciton states is observed in some Eu$^{2+}$ and
Yb$^{2+}$ materials.  The broad, red-shifted emission is often referred
to as ``anomalous'' luminescence~\cite{Do03a}.

Excitonic emission in CaF$_2$ and SrF$_2$ doped with Yb$^{2+}$ has been
the subject of a number of studies~\cite{MoCoPe89,MoCoPe91,PeJoMc07}.
Yb$^{2+}$-doped CaF$_2$, SrF$_2$, and BaF$_2$ form an interesting
series. Though their absorption spectra are similar, with absorption
starting in the UV, the CaF$_2$:Yb$^{2+}$ emission peak is in the
visible, SrF$_2$:Yb$^{2+}$ in the infra-red, and BaF$_2$:Yb$^{2+}$ does
not radiate.

CaF$_2$:Yb$^{2+}$ and SrF$_2$:Yb$^{2+}$ have ground-state electronic
configuration $4f^{14}$. UV excitation can promote one of the $4f$
electrons to a $5d$ orbital, giving the excited configuration
$4f^{13}5d$.  The $5d$ electron rapidly becomes delocalized onto the
next-nearest-neighbor Ca$^{2+}$ or Sr$^{2+}$ ions.  The Yb$^{2+}$ is
then effectively ionized to Yb$^{3+}$, with electronic configuration
$4f^{13}$, i.e.\ one $4f$ \emph{hole}.  This trivalent ion attracts the
F$^-$ nearest neighbors more strongly than a divalent ion, leading to a
large contraction of bond length.  Emission from the exciton states to
the $4f^{14}$ ground state involves a large change in bond length, and
therefore a broad, structureless, red-shifted vibronic emission
band~\cite{Do03a}.  Recent ab-initio calculations have given valuable
insight into the quantum physics of exciton formation~\cite{SaSeBa10a}.
However, the broad bands yield no detailed information and experimental
information on the structure of impurity-trapped excitons has only been
deduced from indirect measurements such as temperature
dependencies~\cite{MoCoPe89,MoCoPe91}, pressure
dependencies~\cite{GrMa08}, and photoconductivity~\cite{PeJoMc07}.

We have recently showed that it is possible to study the internal
structure of trapped excitons in CaF$_2$:Yb$^{2+}$ by using a two-color
photoluminesence-enhancement technique~\cite{ReSeWeBeMeSaDuRe11}.  By
irradiating the crystal with IR radiation after exciting it with UV
radiation we induce transitions between exciton states. Since some of
the excited states have much higher radiative rates than the lowest
exciton state we can detect the excited-state absorption by monitoring
photoluminescence enhancement.  In this work we compare CaF$_2$ and
SrF$_2$ doped with Yb$^{2+}$.

Photoionization is known to occur when these materials are excited with
UV radiation~\cite{MoCoPe89,MoCoPe91,PeJoMc07}, and the mobile electrons
may be trapped by crystal defects, possibly clusters of trivalent
cations~\cite{BeFeRiFlBa98}.  IR radiation is known to be capable of
liberating electron traps~\cite{IzKlViBrGr07}.  Our enhanced
photoluminescence measurement has a component that we interpret as
trap-liberation, and we find that the trap-liberation spectrum is very
similar for both CaF$_2$:Yb$^{2+}$ and SrF$_2$:Yb$^{2+}$.

\section{Experimental}

CaF$_{2}$:Yb$^{2+}$ and SrF$_{2}$:Yb$^{2+}$ crystals were grown using
the vertical Bridgmann technique. The Yb concentration was 0.05\% and
0.1\% respectively and the purity of the starting materials was at least
99.9\%. The UV component of our two-color excitation was made using UV
output of a Quantronix TOPAS traveling-wave optical parametric
amplifier (OPA) providing 3 ps pulses tunable in the 250-400 nm region
of interest in this work at a repetition rate of 1 kHz. Pulsed infrared
excitation was achieved using the Dutch FEL (FELIX) in Nieuwegein. The
optical output of FELIX consists of a 4-6 $\mu$s macropulse at a
repetition rate of 10 Hz, containing micropulses at 25 MHz. FELIX is
continuously tunable from 3 to 250~$\mu$m. The OPA was synchronized to
the FEL and the electronic timing between the two lasers could be
varied. The UV and IR beams were spatially (but not temporally)
overlapped on the sample, held at cryogenic temperatures within an
Oxford instruments \emph{microstat} helium flow cryostat.  Subsequent
UV/visible fluorescence was then detected using a TRIAX 320 spectrometer
equipped with a C31034 photomultiplier.

\section{Results}

\subsection{CaF$_2$:Yb$^{2+}$}

Excitation with 365 nm pulsed UV excitation of CaF$_2$:Yb$^{2+}$ gives
results consistent with previous work~\cite{MoCoPe89,MoCoPe91}, which
reported strongly red-shifted fluorescence having a single-exponential
decay with a lifetime of 15 ms at 4.2~K that reduces at higher
temperatures.

\begin{figure}[tb!]
\begin{center} 
% multiply by 0.9 for preprint, 0.9 for reprint. 
\includegraphics[width=1.0\columnwidth]{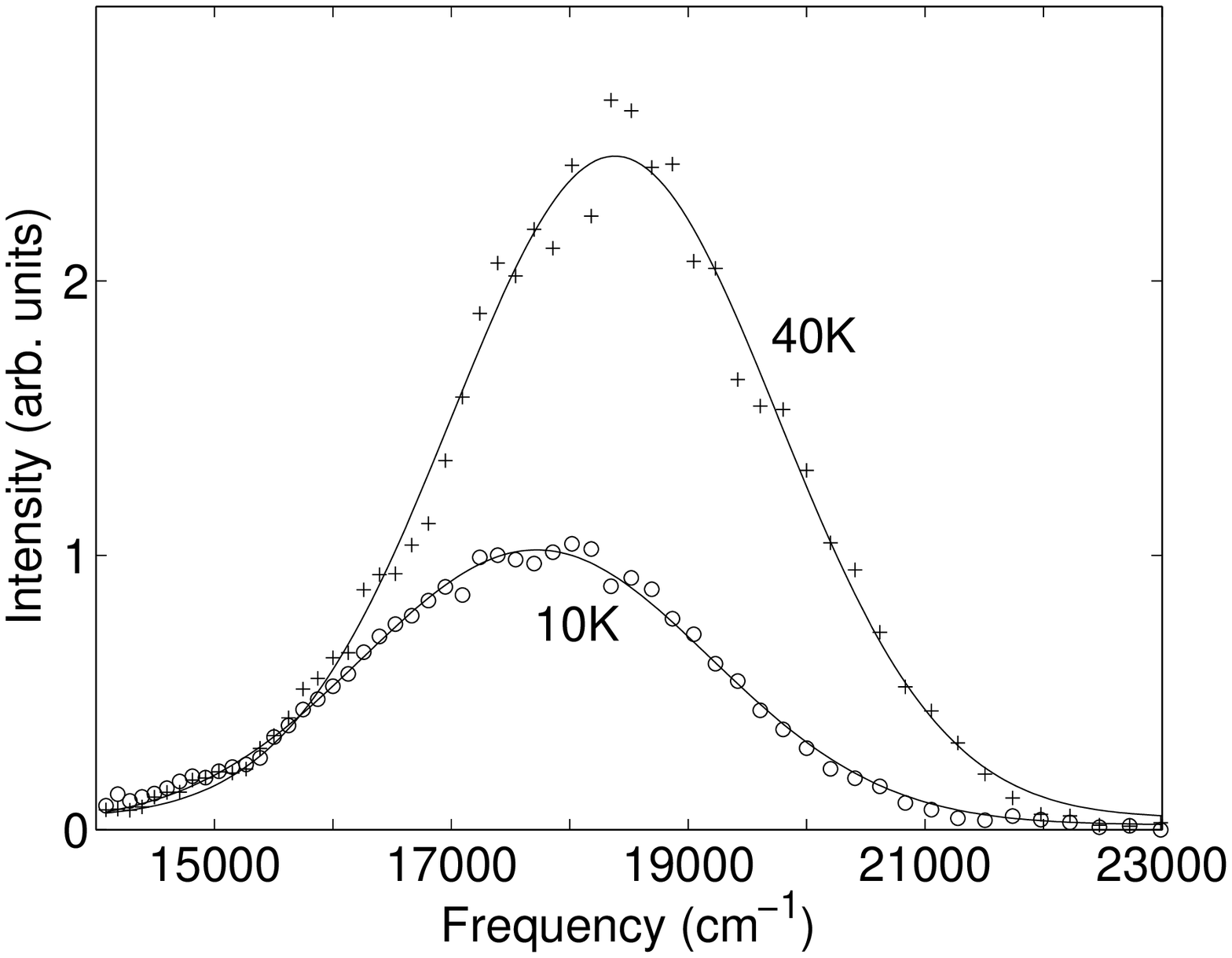}\\  
\caption{\label{fig:emission} Fluorescence from CaF$_2$:Yb$^{2+}$ for
  excitation at 365 nm. Spectra at 10~K and 40~K 900~$\mu$s after
  UV excitation.}
\end{center}
\end{figure}

The fluorescence spectra for CaF$_{2}$:Yb$^{2+}$ at 10~K and 40~K are
shown in Fig.~\ref{fig:emission}.  At 10~K the band center and width
(FWHM) are 17720, 3460~cm$^{-1}$. At 40~K the intensity increases by a
factor of 2.5 and the band center and width are 18380, 3240~cm$^{-1}$.
Previous analysis of the temperature dependence~\cite{MoCoPe89} suggests
that the emission is from two states separated by about 40~cm$^{-1}$,
with very different radiative lifetimes, 15~ms for the lower state and
260~$\mu$s for the upper state.

Bond-length changes may be estimated from the emission bandwidths. The
change in bond length between the lowest exciton state and the ground
state has been calculated to be 0.17~\AA~\cite{MoCoPe89}.  The
calculation is approximate because it uses an effective phonon
frequency, which must be estimated from the phonon
spectrum~\cite{HaWiMaMc73}.  A frequency of 325~cm$^{-1}$ gives 0.17
\AA\ for the lowest exciton state (from our 10~K data) and 0.16 \AA\ for
the first excited state (from our 40~K data). The first excited state
thus has a \emph{longer} bond length, closer to the ground state
$4f^{14}$ bond length.

\begin{figure}[tb!] 
\begin{center} 
% multiply by 0.77 for preprint, 0.9 for reprint.
\includegraphics[width=0.85\columnwidth]{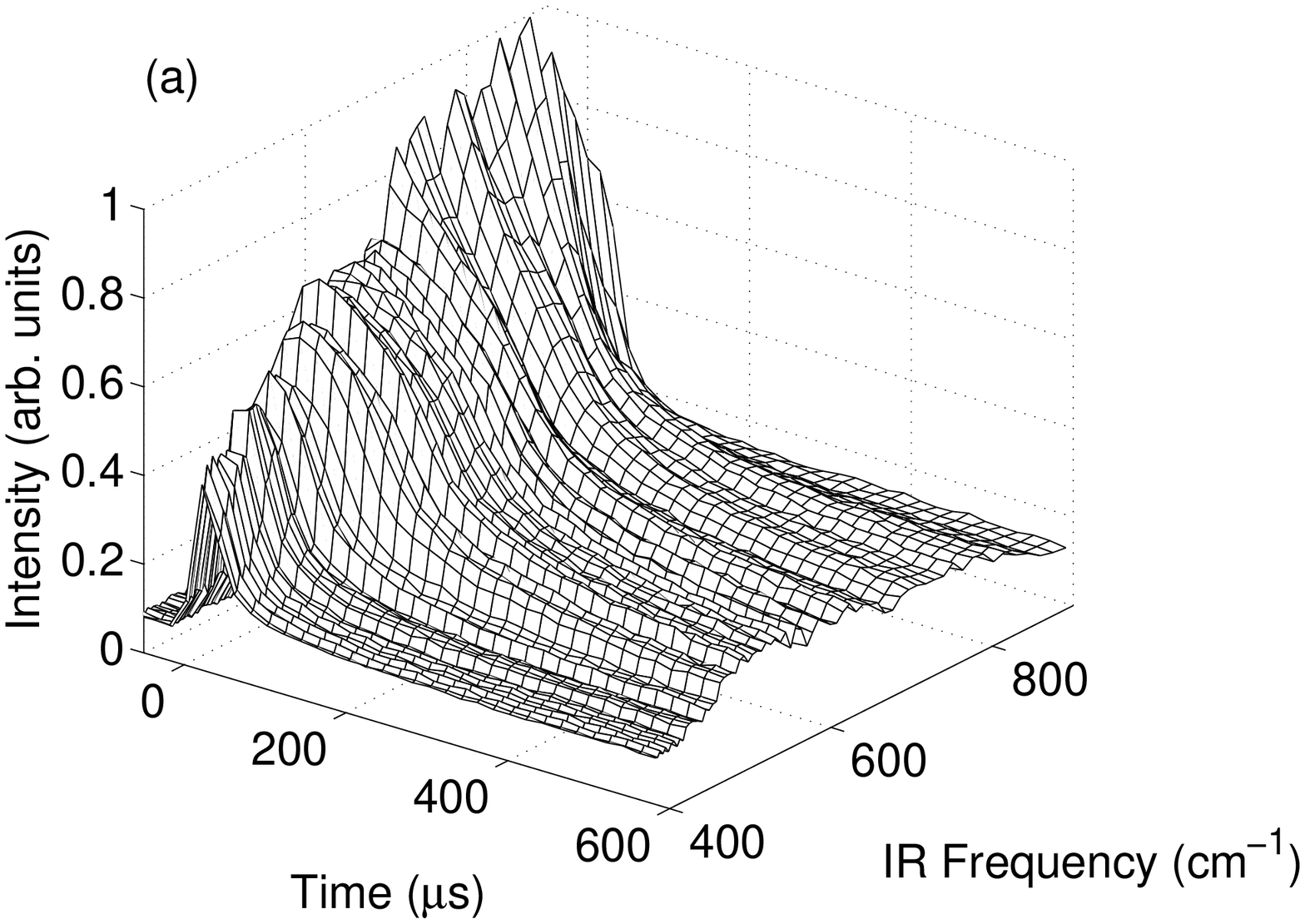}\\

\bigskip

\includegraphics[width=0.85\columnwidth]{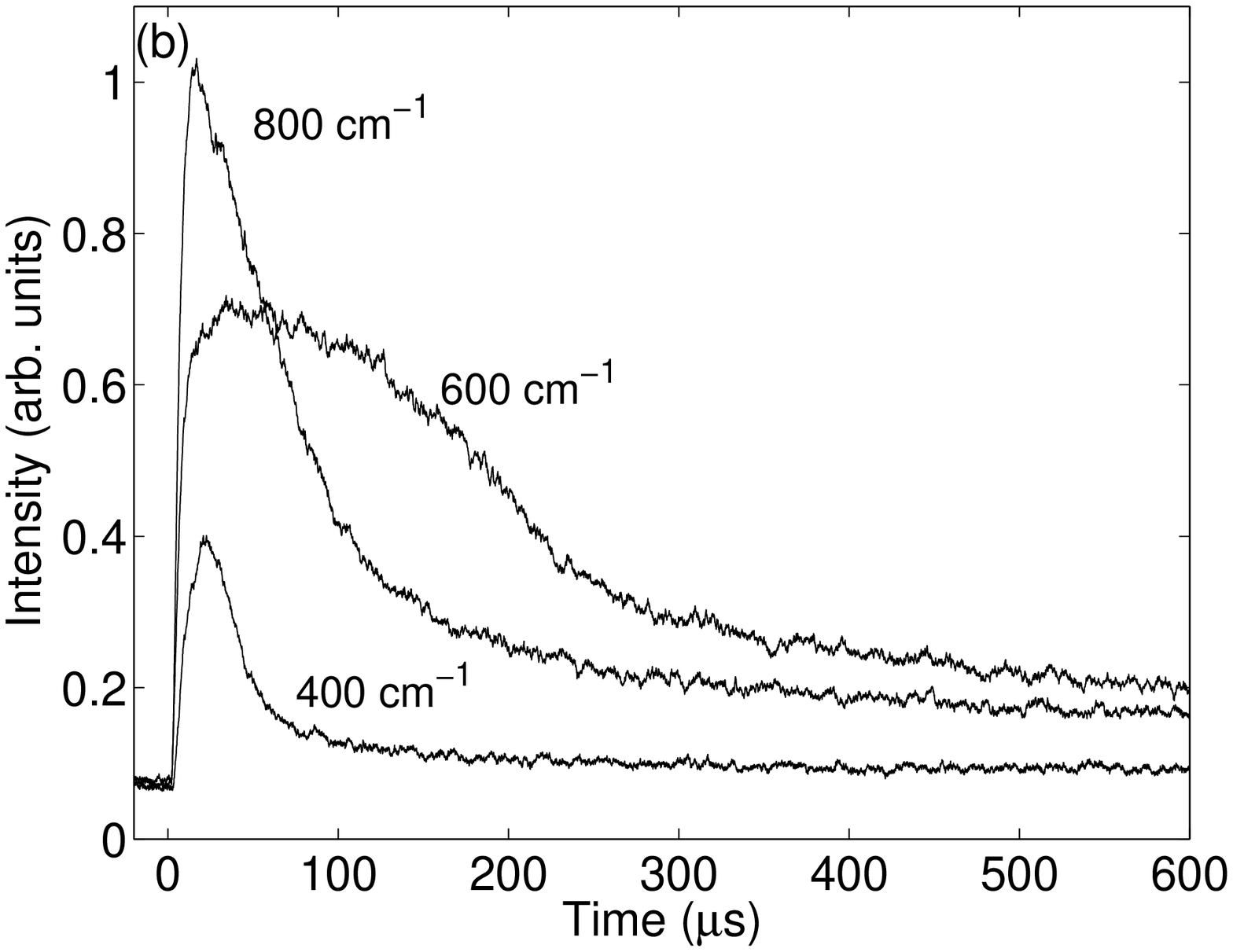}\\
\caption{\label{fig:caf2time} (a) Transient response of
  CaF$_2$:Yb$^{2+}$ excited at IR frequencies of 400--900~cm$^{-1}$
  100~$\mu$s after the UV pulse.  (b) Transient response of
  CaF$_2$:Yb$^{2+}$ excited at IR frequencies of 400, 600, and
  800~cm$^{-1}$. }
\end{center} 
\end{figure}

Two-color transient measurements at 10~K are illustrated in
Fig.~\ref{fig:caf2time}(a), which shows the result of irradiating the
system with IR pulses with frequencies between 400 and 900~cm$^{-1}$,
delayed from the UV excitation by 100~$\mu$s.  The application of
the IR pulse yields significant {\it enhancement} \/of the emission on
short timescales.  This is because we now populate excited excitonic
states that have significantly {\it faster} \/radiative rates.

Fig.~\ref{fig:caf2time}(b) shows the time evolution for frequencies of
400, 600, and 800~cm$^{-1}$. For 400 and 800~cm$^{-1}$ IR excitation the
rise and decay times are similar.  The temporal behavior for
600~cm$^{-1}$ excitation is very different. The decay is clearly not a
single exponential and the rise time is much longer than for 400 and
800~cm$^{-1}$ IR excitation.  Several hundred~$\mu$s after the IR
excitation there is still a significant enhancement, suggesting that
there are many \emph{more} ions radiating than before the IR excitation.

\begin{figure}[tb!] 
\begin{center} 
% multiply by 0.9 for preprint, 0.9 for reprint.
\includegraphics[width=1.0\columnwidth]{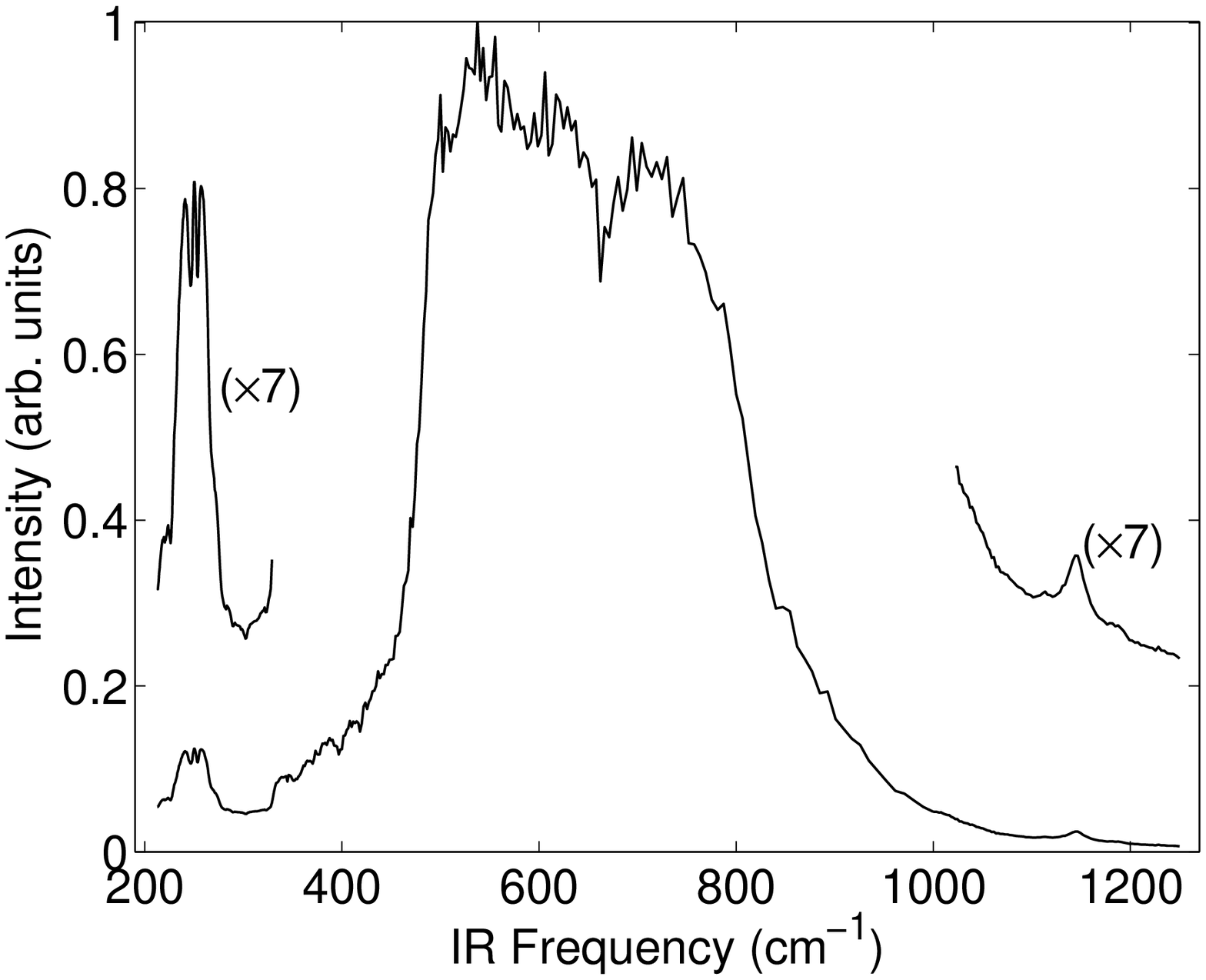}\\
\caption{\label{fig:caf2spectrum} 10~K IR excitation spectrum of
  CaF$_2$:Yb$^{2+}$ deduced by integrating emission enhancement over
  time.  }
\end{center} 
\end{figure}

We may integrate the IR enhancement of the photoluminescence to
construct an IR-induced excitation spectrum. This is shown in
Fig.~\ref{fig:caf2spectrum}. Note that this spectrum has been
constructed from three separate FEL scans.  The spectrum consists of a
broad band centered at 650~cm$^{-1}$ and two sharper peaks at 250
cm$^{-1}$ and 1145~cm$^{-1}$, with linewidths of 35 and 20~cm$^{-1}$
respectively. Dips in the broad band and the lower sharp
peak correlate with atmospheric absorption of the FEL radiation which
can distort the lineshape, despite purging of the IR beam path with dry
N$_{2}$ gas.
 
As discussed above, the temporal dynamics of the IR-induced signal are
strongly dependent on the excitation wavelength. This is clear from
Fig.~\ref{fig:caf2time} which reveals that IR excitation from about 450
to 650~cm$^{-1}$ gives significantly longer decay times than the lower-
and higher-frequency regions.

\begin{figure}[tb!]
\begin{center} 
% multiply by 0.9 for preprint, 0.9 for reprint.
\includegraphics[width=1.0\columnwidth]{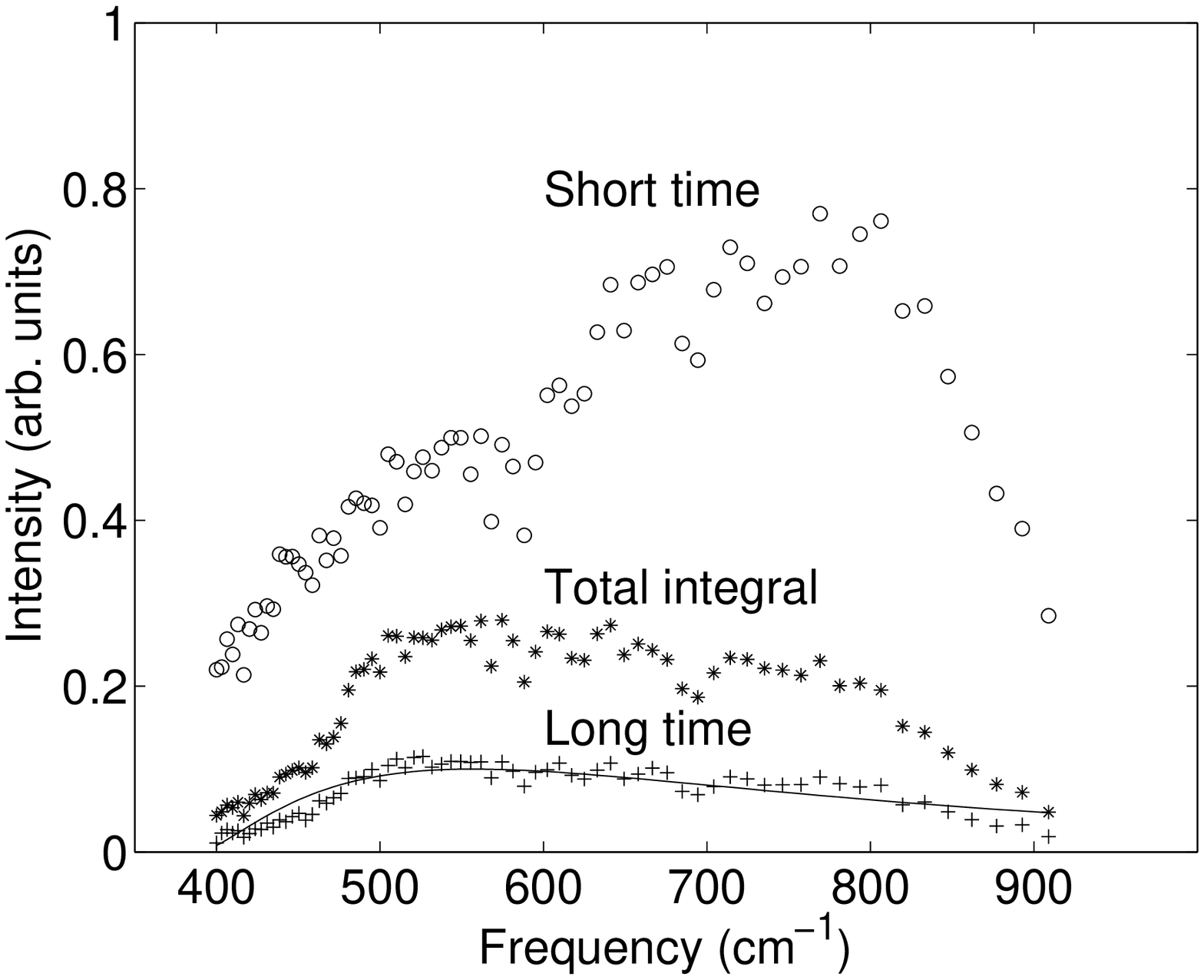}\\
\caption{\label{fig:caf2excitontrap} Time-gated IR excitation spectra of
  CaF$_2$:Yb$^{2+}$ derived from the time-resolved excitation spectrum
  of Fig.~\ref{fig:caf2time}(a).
  The short time spectrum is an average for 0--45~$\mu$s, the total
  integral for 0--750~$\mu$s and the long time for 740--750~$\mu$s. The
  solid curve is for a Coulomb trap with threshold of 390~cm$^{-1}$.  }
\end{center} 
\end{figure}

There are two possible mechanisms for two-color photoluminescence
enhancement in these materials. One is that the IR pulse changes the
exciton to a state with a higher radiative rate. This will be most
apparent at short times. The other is that the IR pulse liberates a
trapped electron, which recombines with a Yb$^{3+}$ ion. This will
result in \emph{more} \/excitons radiating, an enhancement that will
persist at long time scales. Trap liberation should also have a longer
rise time than transition between exciton states. We therefore conclude
that the low-energy IR excitation spectrum as dominated by trap
liberation and the high-energy by exciton transitions.

This interpretation may be made more quantitative by integrating the
data of Fig.~\ref{fig:caf2time}(a) over various time windows. The result
is shown in Fig.~\ref{fig:caf2excitontrap}, where the upper trace is for
short times, the middle an integration over the entire time-scale (as in
Fig.~\ref{fig:caf2spectrum}) and the bottom trace for long times.

The long-time spectrum is presumed to come entirely from trap
liberation.  A phenomenological model of a Coulomb trap having a
threshold of 390~cm$^{-1}$ gives an asymmetric spectrum with a width of
approximately 400~cm$^{-1}$ (\cite{IzKlViBrGr07} Eqn.\ (3)). This curve
is included in Fig.~\ref{fig:caf2excitontrap}, and it gives a reasonable
representation of the long-time spectrum, though it does not fall off at
high energy as fast as the experimental data.

%%%%% Broad exciton band

The excitonic states may be modeled in a similar way to the $4f^{13}5d$
excited states of Yb$^{2+}$~\cite{PaDuTa08,ReHuFrDuXiYi10}, with the
$5d$ electron replaced by a more delocalized orbital.  The broad band
from 650 to 950~cm$^{-1}$ may be attributed to transitions from the
lowest exciton state that involves a change in the orbital of the
delocalized electron, giving a change in bonding and therefore vibronic
broadening. The width of this band is similar to the width of the phonon
spectrum in CaF$_2$~\cite{HaWiMaMc73}, which implies that the
bond-length change for the transition is small, similar to the change in
bond-length between the lowest two exciton states calculated above (0.01
\AA).  An accurate calculation of the position and width of this band
should be possible with a detailed ab-initio approach, as in
Ref.~\cite{SaSeBa10a}.

%%%%% Analysis of Sharp lines. 
 
The sharp excitation features observed at 250 and 1145~cm$^{-1}$ cannot
involve a change in bonding. We therefore assign them to changes in the
wavefunction of the localized $4f$ hole or the relative spin of the $4f$
hole and delocalized electron. The former is a ``crystal field''
interaction and the latter is associated with an exchange Coulomb
interaction. The calculations of Ref.~\cite{SaSeBa10a} suggest that the
excitons involve a linear combination of $5d$ and $6s$ orbitals with
totally symmetric ($s$) character.  Detailed calculations for
CaF$_2$:Yb$^{3+}$ are not available.  In Ref.~\cite{ReSeWeBeMeSaDuRe11}
we have modeled the sharp lines with a simple semi-empirical model by
constructing a ``crystal field'' Hamiltonian for an $s$ electron and a
$4f$ hole in a cubic crystal field. The 250~cm$^{-1}$ splitting is
simply the crystal-field splitting of the $4f^{13}$ configuration of a
Yb$^{3+}$ ion in a cubic site, whereas the 1145~cm$^{-1}$ splitting also
involves the exchange interaction between the $4f$ and delocalized
electrons.

\subsection{ SrF$_2$:Yb$^{2+}$}

We now turn to SrF$_2$:Yb$^{2+}$. The red shift of the exciton emission
in this material is very large, with the peak emission in the near
infra-red, and non-radiative processes are much more rapid than in
CaF$_2$:Yb$^{2+}$~\cite{MoCoPe89,MoCoPe91}. Since the proportion of ions
that decay radiatively is lower, and the detection was not optimized for
IR radiation, the the noise levels for measurements in this material are
much higher.

\begin{figure}[tb!]
\begin{center}
% multiply by 0.77 for preprint, 0.9 for reprint.
\includegraphics[width=0.85\columnwidth]{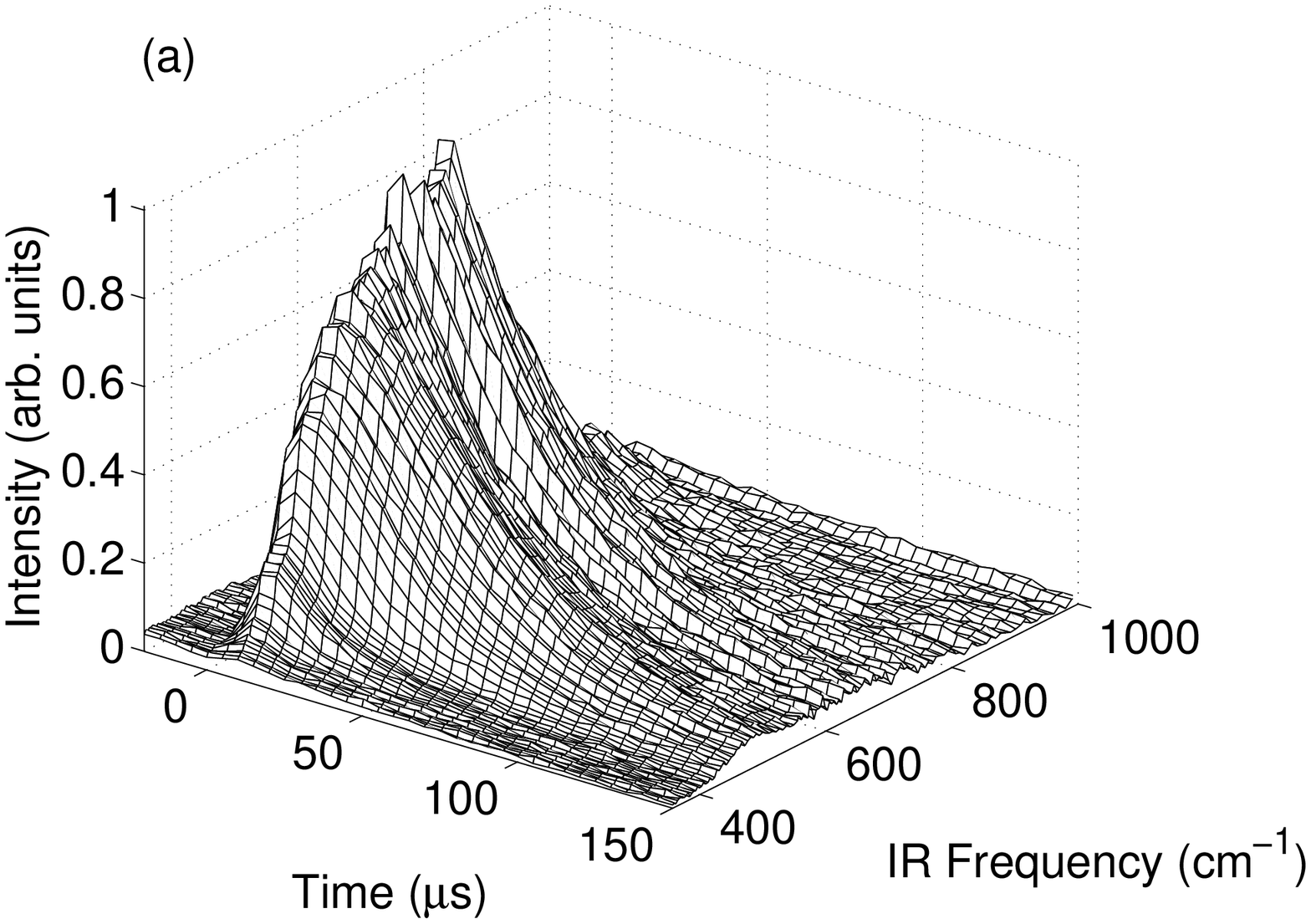}\\

\bigskip

\includegraphics[width=0.85\columnwidth]{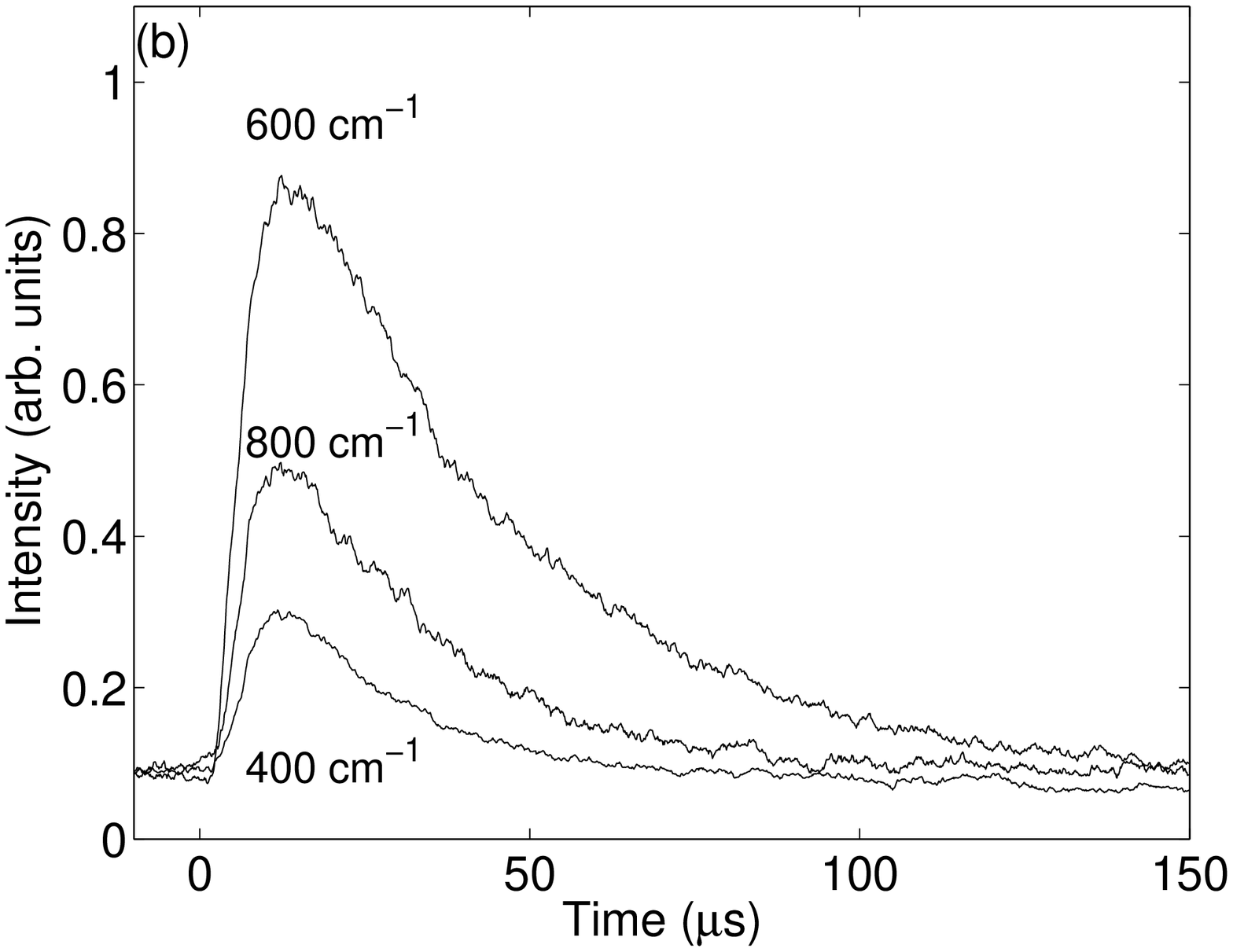}\\
\caption{\label{fig:srf2time} (a) Transient response of
  SrF$_2$:Yb$^{2+}$ excited at IR frequencies of 350--1000~cm$^{-1}$
   100~$\mu$s after the UV pulse.  (b) Transient response of
  SrF$_2$:Yb$^{2+}$ excited at IR frequencies of 400, 600, and
  800~cm$^{-1}$. }
\end{center}
\end{figure}

Fig.~\ref{fig:srf2time}(a) shows the result of irradiating the system
with IR pulses with frequencies from 350 to 1000~cm$^{-1}$, delayed
from the UV excitation by 100~$\mu$s. Fig.~\ref{fig:srf2time}(b) shows
the trace for frequencies of 400, 600, and 800~cm$^{-1}$. In contrast to
the CaF$_2$:Yb$^{2+}$ case, there is much less variation across the
spectrum, but there an increase in decay time in the middle of the broad
band.

\begin{figure}[tb!]
\begin{center} 
% multiply by 0.9 for preprint, 0.9 for reprint.
\includegraphics[width=1.0\columnwidth]{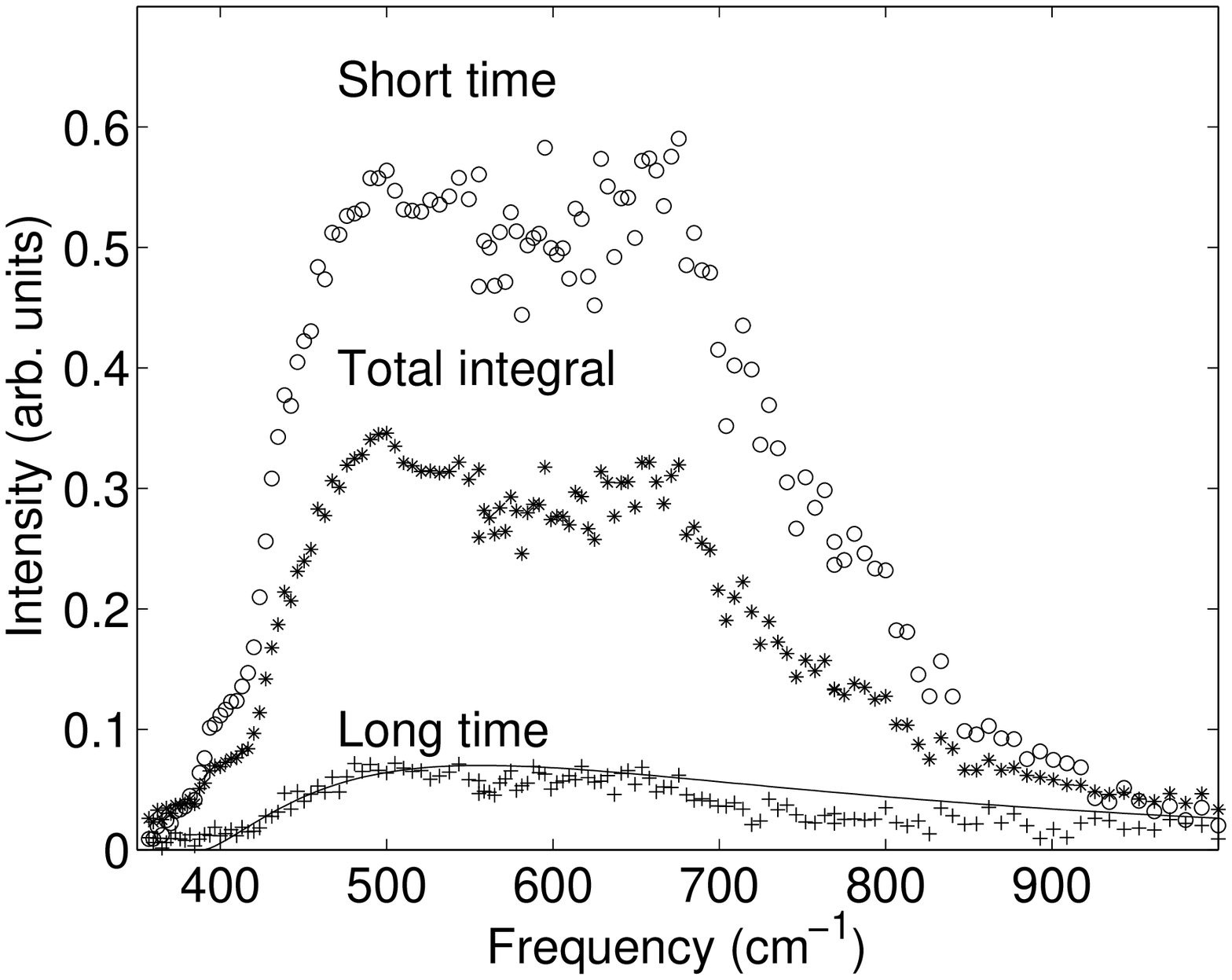}\\  
\caption{\label{fig:srf2excitontrap} Time-gated IR excitation spectra of
  SrF$_2$:Yb$^{2+}$ derived from the time-resolved excitation spectrum
  of Fig.~\ref{fig:srf2time}(a).  The top trace is an average for
  0--45~$\mu$s, the middle for 0--150~$\mu$s and the bottom trace for
  140--150~$\mu$s. The solid curve is for a Coulomb trap with threshold
  of 390~cm$^{-1}$.  }
\end{center}
\end{figure}

Fig.~\ref{fig:srf2excitontrap} shows spectra derived from the
time-resolved excitation spectrum of Fig.~\ref{fig:caf2time}(a). The top
trace is for short times, the middle an integration over the entire
time-scale, and the bottom trace for long times. The theoretical curve
for a Coulomb trap matches the long-time spectrum quite well.

Comparison of the short and long time spectra suggest that there is an
exciton peak at about 670~cm$^{-1}$, and we conclude that in the larger
SrF$_2$ lattice the energy difference between exciton levels is smaller
than in SrF$_2$.  We have not yet observed any sharp lines in the
SrF$_2$:Yb$^{2+}$ IR excitation spectrum. However, the noise levels make
detection difficult in the region where a high-energy peak might be
expected (1000---1100 cm$^{-1}$).  We expect that measurements at lower
IR energies will reveal similar $4f^{13}$ crystal-field splitting as for
CaF$_2$.

\subsection{Comparison of trap liberation in CaF$_2$:Yb$^{2+}$ and SrF$_2$:Yb$^{2+}$}

\begin{figure}[tb!]
\begin{center} 
% multiply by 0.9 for preprint, 0.9 for reprint.
\includegraphics[width=1.0\columnwidth]{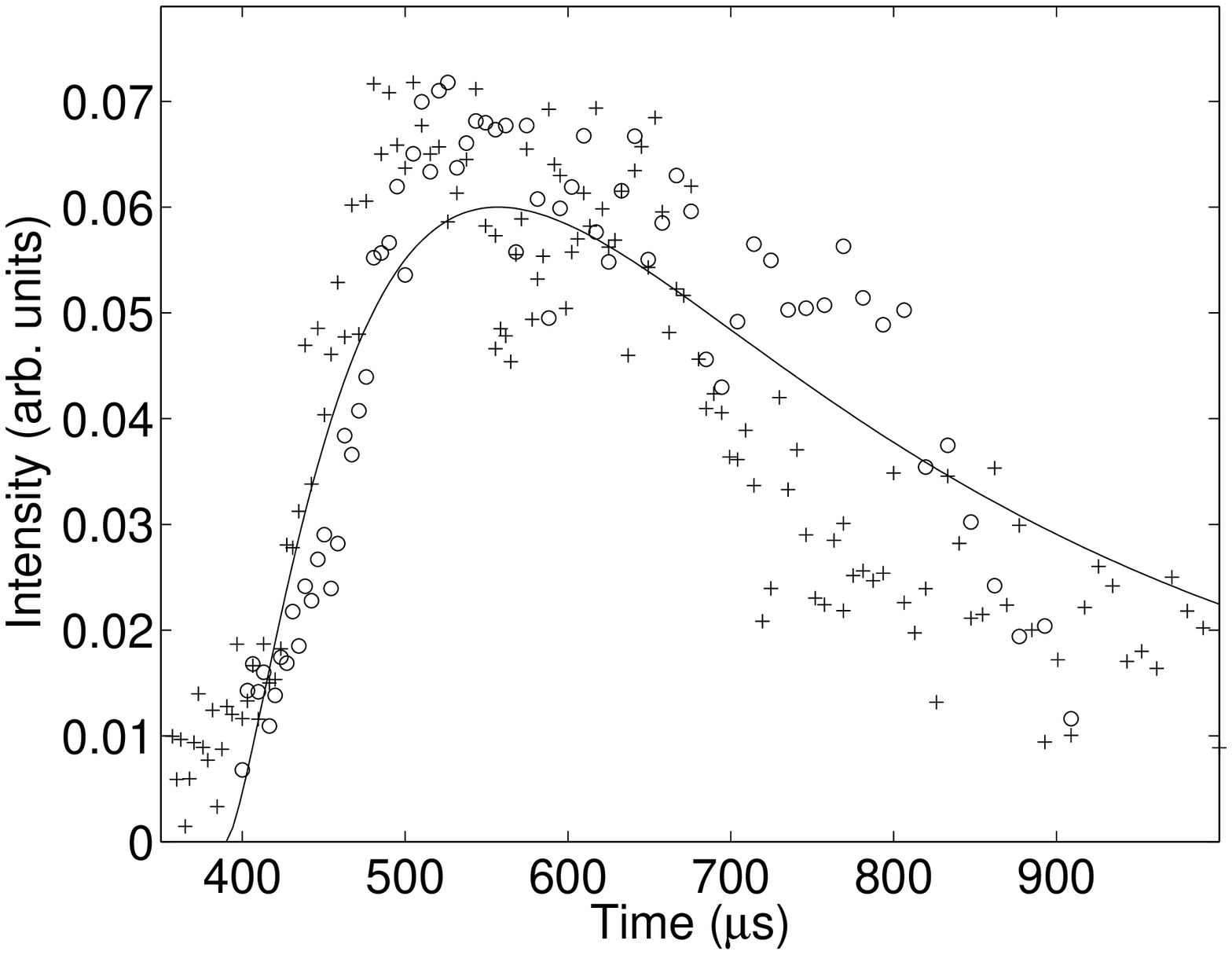}\\
\caption{\label{fig:casrtrap} Comparison of long-time IR excitation
  spectra of CaF$_2$:Yb$^{2+}$ ($\circ$) and SrF$_2$:Yb$^{2+}$ ($+$)  from
  Figs.~\ref{fig:caf2excitontrap} and \ref{fig:srf2excitontrap}.  The
  solid curve is a Coulomb trap model with threshold 390~cm$^{-1}$.  }
\end{center} 
\end{figure}

The results CaF$_2$:Yb$^{2+}$ and SrF$_2$:Yb$^{2+}$ presented above
suggest that the trap-liberation spectrum for both materials is very
similar. This is further illustrated in Fig.~\ref{fig:casrtrap}, which
shows the long-time two-color spectra for both CaF$_2$:Yb$^{2+}$ and
SrF$_2$:Yb$^{2+}$, along with the Coulomb-trap model curve. Unlike the
exciton spectra there is apparently little variation of the
trap-liberation spectra with lattice constant. This would be expected if
the traps are clusters of trivalent ions~\cite{BeFeRiFlBa98}.

\section{Conclusions}

A two-color selective fluorescence-enhancement technique has been used
to investigate CaF$_2$:Yb$^{2+}$ and SrF$_2$:Yb$^{2+}$.  We have
demonstrated that it is possible to use this technique to probe both the
internal structure of exciton states and trap-liberation processes.

The spectra show both broad-band and sharp-line absorptions
\emph{within} the exciton, which will provide stringent tests of
ab-initio calculations such as~\cite{SaSeBa10a}. They also provide
information about the dynamics of relaxation within the exciton that are
not available from one-color measurements. 

Trap-liberation processes are also probed by our measurements. The
spectra are in reasonable agreement with a simple Coulomb-trap model.
The trap spectra in CaF$_2$:Yb$^{2+}$ and SrF$_2$:Yb$^{2+}$ are similar,
which is to be expected if the electron traps are clusters of cations.

\section*{Acknowledgements}

This work was supported by the Marsden fund of the Royal Society of New
Zealand via grant 09-UOC-080. We thank the Dutch FOM organization for
providing FELIX beamtime and thank the FELIX staff for their assistance.
Mr P. S. Senanayake acknowledges the support of the University of
Canterbury via a PhD studentship.



%% References


\bibliographystyle{model1a-num-names}
%\bibliography{exciton}






% \end{thebibliography}

\end{document}